\documentclass[12pt]{article}
\pdfoutput=1
\usepackage{jheppub}
\usepackage[utf8]{inputenc}
\usepackage{verbatim}
\usepackage{amsmath}
\usepackage{amssymb}
\usepackage{amsthm}
\usepackage{slashed}
\usepackage{amsfonts}

\newcommand{\be}{\begin{equation}}
\newcommand{\ee}{\end{equation}}
\newcommand{\bfig}{\begin{figure}\begin{center}}
\newcommand{\efig}{\end{center}\end{figure}}
\newcommand{\bi}{\begin{itemize}}
\newcommand{\ei}{\end{itemize}}

\theoremstyle{definition}

\begin{document}
\title{Simple holographic models of black hole evaporation}
\author{Chris Akers,}
\author{Netta Engelhardt,}
\author{and Daniel Harlow}
\affiliation{Center for Theoretical Physics\\ Massachusetts Institute of Technology, Cambridge, MA 02139, USA}
\emailAdd{cakers@mit.edu,engeln@mit.edu,harlow@mit.edu}
\abstract{Several recent papers have shown a close relationship between entanglement wedge reconstruction and the unitarity of black hole evaporation in AdS/CFT.  The analysis of these papers however has a rather puzzling feature: all calculations are done using bulk dynamics which are essentially those Hawking used to predict information loss, but applying ideas from entanglement wedge reconstruction seems to suggest a Page curve which is consistent with information conservation.  Why should two different calculations in the same model give different answers for the Page curve?  

In this note we present a new pair of models which clarify this situation.  Our first model gives a holographic illustration of unitary black hole evaporation, in which the analogue of the Hawking radiation purifies itself as expected, and this purification is reproduced by the entanglement wedge analysis.  Moreover a smooth black hole interior persists until the last stages the evaporation process.  Our second model gives an alternative holographic interpretation of the situation where the bulk evolution leads to information loss: unlike in the models proposed so far, this bulk information loss is correctly reproduced by the entanglement wedge analysis.  This serves as an illustration that quantum extremal surfaces are in some sense kinematic: the time-dependence of the entropy they compute depends on the choice of bulk dynamics. In both models no bulk quantum corrections need to be considered: classical extremal surfaces are enough to do the job.   We argue that our first model is the one which gives the right analogy for what actually happens to evaporating black holes, but we also emphasize that any complete resolution of the information problem will require an understanding of non-perturbative bulk dynamics.}   
\maketitle
\section{Introduction and review}
The black hole information problem is one of the great challenges facing physics today: it is clearly telling us something deep about the combination of quantum mechanics and gravity, and if only we can learn to think about it properly we may well see the way forward on many related fronts \cite{Hawking:1976ra,Harlow:2014yka,Marolf:2017jkr}.  Recently a new approach to the black hole information problem has been introduced \cite{Almheiri:2019psf,Penington:2019npb,Almheiri:2019hni}: we begin with a holographic CFT in a state which is dual to a ``large'' AdS black hole and then couple it to an exterior ``auxiliary system'', allowing the black hole to evaporate (see \cite{Rocha:2008fe} for an earlier discussion of this idea).  The evolving system may then be studied using the ``quantum extremal surface'' proposal of \cite{Engelhardt:2014gca}, using the tool of entanglement wedge reconstruction \cite{Czech:2012bh,Wall:2012uf,Headrick:2014cta,Jafferis:2015del,Dong:2016eik} to see when the auxiliary system begins to contain information about the black hole interior \cite{Almheiri:2019psf,Penington:2019npb,Almheiri:2019hni}.

The main technical result of \cite{Almheiri:2019psf,Penington:2019npb} was the discovery of an previously-unknown bulk quantum extremal surface, which at late times lies just inside the event horizon of a black hole evaporating via its coupling to an auxiliary system.  Since the area of this horizon is decreasing as the evaporation proceeds, the area (plus quantum corrections) of this quantum extremal surface is also decreasing.  Therefore if this surface can be argued to be the correct quantum extremal surface for computing the von Neumann entropy of the auxiliary system, its decreasing area (plus quantum corrections) maps directly to the decreasing entropy in the latter piece of the Page curve, as expected for unitary evaporation \cite{Almheiri:2019psf,Penington:2019npb}.  In \cite{Almheiri:2019psf} this surface was found explicitly for the special case of two-dimensional conformal matter coupled to Jackiw-Teitelboim gravity, while in \cite{Penington:2019npb} it was constructed somewhat less explicitly in higher dimensions assuming spherical symmetry.  In both cases, the new quantum extremal surface only dominates the entropy calculation a scrambling time after the Page time; the analyses of~\cite{Almheiri:2019psf,Penington:2019npb} show that the surface evolves in a spacelike direction, providing a bulk realization of the Hayden-Preskill protocol so long as this new quantum extremal surface correctly defines the quantum corrected entanglement wedge. Ref. \cite{Almheiri:2019hni} then studied a special case of the results of \cite{Almheiri:2019psf} where the conformal matter is taken to have a semiclassical holographic dual, leading to an effective \textit{three-dimensional} description of the entropy of the conformal matter fields.  

There is something puzzling about the results of \cite{Almheiri:2019psf,Penington:2019npb,Almheiri:2019hni}.  This is that all calculations were done within the framework of perturbative bulk gravity coupled to matter fields; precisely the regime where Hawking's calculations show information loss.  This is not necessarily a logical contradiction with the Page curve they found using quantum extremal surfaces: perturbative bulk quantum gravity is not expected to be holographic by itself, so there is not necessarily any useful interpretation for its quantum extremal surfaces.  On the other hand, the fact that this ``wrong'' calculation gives the ``right'' answer seems likely to be telling us something very interesting about the information problem; we just need to interpret it carefully.  In particular we need to be unambiguous about what any particular model is telling us about the following question: \textit{if we, living in a gravitational system, were to collect the full Hawking radiation from an evaporated black hole and carefully study it using mirrors, beam splitters, etc., would we find a pure state or a mixed state?}  In \cite{Almheiri:2019psf,Penington:2019npb,Almheiri:2019hni} no firm answer was given to this question, and in particular in \cite{Almheiri:2019hni} it was suggested that the answer can depend on ``how we obtained the state''.    We are not satisfied with such ambiguity, and removing it is the goal of this paper.  To that end we present a pair of holographic models of black hole evaporation, for one of which the answer to the italicized question is ``yes'' while for the other it is ``no''.\footnote{It is ok for \textit{different} models to give different answers to this question: what is not ok is for the \textit{same} model to give us different answers!}  Moreover in both cases the answer is the same regardless of whether we compute it using the quantum extremal surface prescription or evaluate it directly in the CFT.  After presenting our models we will revisit the results of \cite{Almheiri:2019psf,Penington:2019npb,Almheiri:2019hni}, emphasizing that the tension between the ``bulk'' and ``boundary'' Page curves they found is only present if one is not clear about the relationship between the bulk theory and its holographic dual.   In our models this relationship is explicit, and there is no ambiguity in the Page curve.

\section{Multi-boundary wormholes}
In the models of \cite{Almheiri:2019psf,Penington:2019npb,Almheiri:2019hni}, the entropy of bulk matter fields was essential in constructing the new quantum extremal surface which allows the Page curve to decrease.  In \cite{Harlow:2016vwg} it was pointed out that black holes in AdS/CFT have the interesting property that in the holographic entropy formula 
\be
S_{CFT}=\frac{\mathrm{Area}(\gamma)}{4G}+S_{bulk},
\ee
where $\gamma$ is the quantum extremal surface of smallest generalized entropy \cite{Ryu:2006bv, Hubeny:2007xt, Faulkner:2013ana, Engelhardt:2014gca}, there is some freedom in whether we view their microstates as contributing to the area term or the bulk entropy term. This observation was recently developed substantially in \cite{Hayden:2018khn,Akers:2019wxj}.  Here we will take advantage of this flexibility to introduce models where we replace the Hawking radiation particles by large black holes, whose entropy we will always treat as contributing to the area rather than the bulk entropy.  This will then enable a completely geometric picture of the kind of situation studied in \cite{Almheiri:2019psf,Penington:2019npb,Almheiri:2019hni}.\footnote{In \cite{Almheiri:2019hni} an unconventional geometric picture in one dimension higher was given in the special case where all bulk matter is holographic. What happens in our models is similar in spirit, in the sense that in both cases no difficult bulk entropy calculations are necessary, but our setups are simpler since everything is geometric in the conventional sense of AdS/CFT.} Because our models require no quantum corrections, we are therefore able to use HRT surfaces (that is, minimal classically extremal surfaces) rather than having to consider more technically-involved quantum extremal surfaces.

\bfig
\includegraphics[height=5cm]{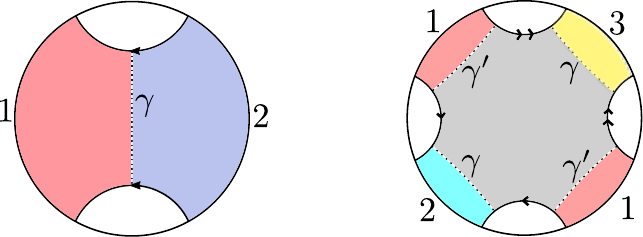}
\caption{Constructing wormhole spatial geometries by quotients of the Poincar\'e disk.  On the left we remove two semicircles to and identify to produce a Cauchy slice of the BTZ wormhole, with the bifurcate horizon being the dotted line marked $\gamma$.  The topology is that of a cylinder.  On the right we remove four semicircles to construct a Cauchy slice of a three-exit wormhole with the topology of a pair of pants.  The exits are labeled $1,2,3$, the shared interior is shaded grey, and a pair of extremal surfaces $\gamma$ and $\gamma'$ which are both homologous to exit $1$ are marked with dotted lines.  In our unitary model below, $\gamma'$ will be the HRT surface at late times and $\gamma$ will be the HRT surface at early times.}\label{wormholefig}
\efig
The basic tool we will use for our construction is multi-boundary wormholes.  These have a long history in $AdS_3/CFT_2$ \cite{Krasnov:2000zq,Skenderis:2009ju,Balasubramanian:2014hda}, where they are simpler to understand since, due to the absence of gravitational waves, they can all be realized as discrete quotients of $AdS_3$.  Here we will only quote a few results which we will need in constructing our models; for a fairly compact recent review see appendix H of \cite{Harlow:2018tng}.  Multi-boundary wormholes arise from CFT states which are prepared by evaluating the Euclidean CFT path integral on a cut Riemann surface \cite{Krasnov:2000zq}.  In favorable regions of the moduli space of the Riemann surface, we obtain a bulk geometry where arbitrary numbers of copies of the boundary CFT are connected through a shared interior.  For our purposes the key point is that the topology and geometry of the time reflection-symmetric Cauchy slice of such a wormhole can be obtained by a quotient of the two-dimensional Poincar\'e disk, obtained by removing a collection of semicircles from the geometry at the boundary and then identifying their arcs.  We illustrate this construction for the thermofield double state and a simple three-boundary wormhole in figure \ref{wormholefig}.  The key points to appreciate from this construction are:
\bi
\item The minimal length geodesics between the semicircles, shown as dotted lines in figure \ref{wormholefig}, are also geodesics in the full Lorentzian continuation of this geometry.  They are therefore candidate HRT surfaces.  
\item For three or more exits, the moduli of the Riemann surface (which here are just the locations and sizes of the semicircles modulo $SL(2,\mathbb{R})$ transformations) can be adjusted to independently change the lengths of the dotted geodesics.  In particular we can take all of them to have a length which is large compared to the AdS radius, and we can choose one of them to have a length which is much longer than the rest.
\ei
An important question which we do not address in this paper is when these wormhole geometries are the dominant ones in the bulk path integral.  For example in the two-boundary case, the BTZ wormhole is dominant only for high enough temperature.  We will simply assume that, perhaps after some additional tuning of the moduli, the wormholes we discuss are dominant.  If this is not achievable, then our state preparation will need to be modified to include a projection that removes whatever geometry is more dominant.

\section{A model for information conservation}

\bfig
\includegraphics[height=4.3cm]{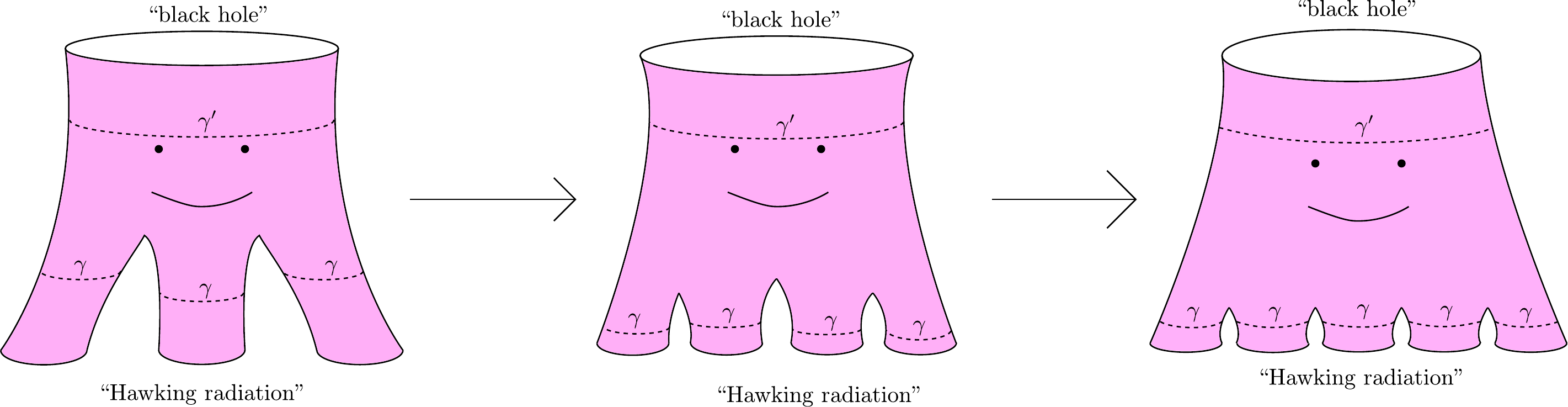}
\caption{Two steps of the ``black hole evaporation'' process for our first model.  The surfaces $\gamma$ and $\gamma'$ give two possible HRT surfaces homologous either to the ``Hawking radiation'' exits or the ``black hole'' exit.  As the octopus grows more and more legs, the length of the ``ankle bracelets'' $\gamma$ increases and the length of the ``headband'' $\gamma'$ decreases: the crossover point where the headband becomes shorter is the ``Page time'' of the model.}\label{octopi} 
\efig
We now present our model for unitary black hole evaporation.  The basic idea is that we want to start with a single pure-state black hole whose size is many times that of the $AdS$ length, and then allow it to evaporate by emission of $AdS$-sized black holes which are parametrically smaller than the one we started with.  To make the analysis simple, we will arrange for this emission to place each such black hole into its own separate asymptotically-AdS spacetime.  As the evaporation proceeds, these asymptotically-AdS regions will be connected by a wormhole with an increasing number of exits.  Holographically we can describe this as follows: we begin with a pure state in the tensor product of many copies of some holographic $CFT_2$, which is a product between a highly-excited pure state in one of the copies and the vacuum in all of the others.  We then couple the CFTs together using a Hamiltonian which is carefully designed to evolve this state to a sequence of others which are prepared by cutting a CFT Riemann surface of higher and higher genus.  This is easier to visualize than to describe, we give an illustration of how the spatial geometry changes over a pair of time steps in figure \ref{octopi}.\footnote{This process does not quite work for the first two steps of the radiation, since it is only once we have at least three exits that we can adjust the sizes of the exits independently.  To fix this we could add some handles behind the horizon to the initial pure state with one exit, as in figure \ref{infolossworm} below, but to avoid complicating the figure here we will instead just begin our evaporation at three exits: we begin with an ``almost pure'' black hole which has radiated only two ``Hawking quanta'' so far.} To avoid confusion we emphasize that this dynamics constitutes a modification of usual bulk dynamics for this system: the topology changes, the area of the black hole horizon decreases with time, and there are direct couplings between the boundary CFTs.  Nonetheless at any fixed time the state is one which can be given a standard holographic interpretation, and in particular for which the boundary entropies can be computed using the standard quantum extremal surface technology.

For black holes in $AdS_3$ the energy-entropy relation is 
\be
S=2\pi \sqrt{\frac{cE}{3}}.
\ee
For simplicity we will require the dynamics to preserve the total energy, and also to have the property that all of the ``Hawking radiation'' exits have the same length $\ell$.  If we denote the initial horizon length by $L_0$, then as a function of (discrete) time $n$ we have
\be\label{Lgp}
L(\gamma')=\sqrt{L_0^2-n\ell^2}.
\ee
Here $\gamma'$ is one of the two candidate HRT surfaces for the CFT containing the original black hole, as shown in figures \ref{wormholefig}, \ref{octopi} (it is the one which is closer to the exterior of that black hole).  $\gamma$ is the other candidate HRT surface, it is the one which is closer to the ``Hawking radiation'' exterior.  Its length is
\be\label{Lg}
L(\gamma)=n\ell.
\ee

\bfig
\includegraphics[height=4cm]{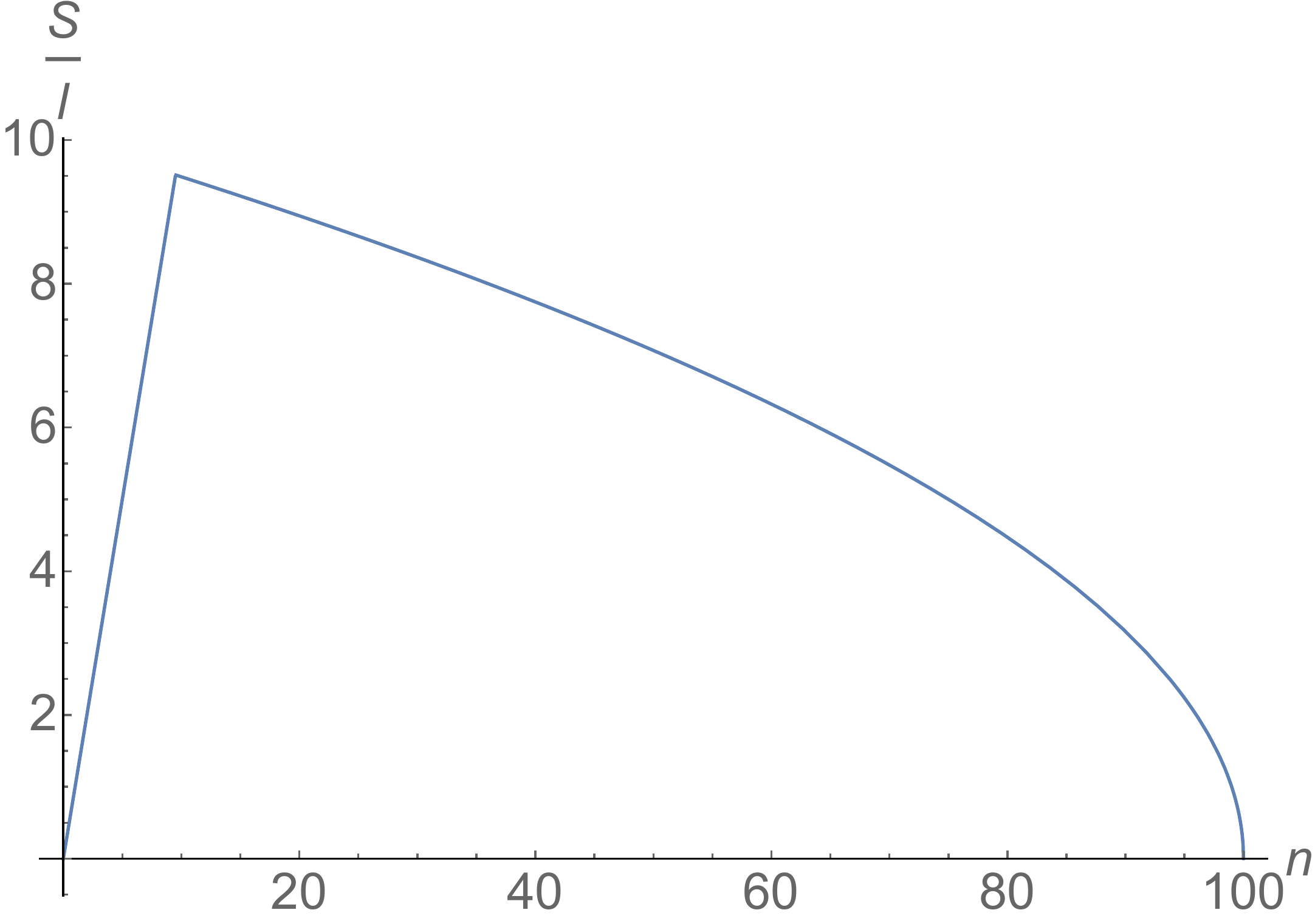}
\caption{The Page curve for our unitary model, computed using the area of the HRT surface (we've taken $L_0/\ell=10$ and $4G=1$, and plotted $\frac{S_R}{\ell}$).}\label{pageplot}
\efig
Since we are working in a situation where $L_0\gg \ell$, in the beginning we will have $L(\gamma)\ll L(\gamma')$ so $\gamma$ will be the HRT surface of either the black hole CFT or the radiation CFTs.  The entanglement wedge of the black hole CFT will contain the full interior piece of this Cauchy slice of the wormhole.  The entanglement wedge of the Hawking radiation will contain only the ``legs'' of the octopus.  As the evaporation proceeds however, we  eventually get to a time where $n\sim \frac{L_0}{\ell}$.  At this point, the ``Page time'', we have $L(\gamma)\approx L(\gamma')$.  Afterwards it is $\gamma'$ which is now the HRT surface, and the interior piece of the slice moves over to being in the entanglement wedge of the radiation.  Thus in our model $\gamma'$ is the analogue of the new quantum extremal surface found in \cite{Almheiri:2019psf,Penington:2019npb,Almheiri:2019hni}.  Since the area of $\gamma'$ is decreasing, we therefore reproduce the qualitative features of the Page curve; we give a plot in figure \ref{pageplot}.   Moreover after the Page time, information about objects we throw into remaining black hole becomes accessible from the Hawking radiation via entanglement wedge reconstruction \cite{Czech:2012bh,Wall:2012uf,Headrick:2014cta,Jafferis:2015del,Dong:2016eik}, as predicted in \cite{Hayden:2007cs}.

We emphasize that although we computed this Page curve using the area of the HRT surface, we could also have computed it directly using the Riemann surface preparation of these states in the CFT: by the argument of \cite{Lewkowycz:2013nqa} it would have to give the same answer.  Moreover the bulk picture of our Hawking radiation just consists of a set of black holes which are in the entangled state described by the CFT state, so the ``bulk'' state of the Hawking radiation is pure.  We also emphasize that there is a smooth semiclassical interior for this geometry well past the Page time; this is an illustration of the ``geometry from entanglement'' philosophy of \cite{VanRaamsdonk:2010pw,VanRaamsdonk:2013sza,Maldacena:2013xja}.  Indeed our late time geometry can be thought of as a classical version of the ``quantum octopus'' from figure 13 of \cite{Maldacena:2013xja}.

This model may seem to be far from the Hawking evaporation process.  For one thing our big black hole radiates into other black holes instead of soft quanta, and for another after each step the state is time-symmetric with respect to the usual evolution.  We make three comments about this:
\bi
\item The various ``firewall'' paradoxes of \cite{Mathur:2009hf,Almheiri:2012rt,Almheiri:2013hfa,Marolf:2013dba} do not arise from the detailed properties of the Hawking process, they instead are proposed contradictions about the properties of the state at a fixed time.  Any evaporation model which describes a smooth interior at late times must grapple with them.  
\item The time-symmetry at each state of our model is just a simplifying assumption which makes it easy to find the extremal surfaces.  Essentially it arises because we have chosen to ``cancel'' the natural Lorentzian evolution going on between the evaporation steps, which would tend to make the shared interior grow.  As this evolution is a product of unitary operators on the different boundaries, it doesn't effect the Page curve and we are thus free to remove it. 
\item Once we accept the possibility of wormholes connecting small quantum systems, we see no objection to realizing the qualitative features of our model in genuine Hawking evaporation into soft quanta.
\ei

\bfig
\includegraphics[height=5cm]{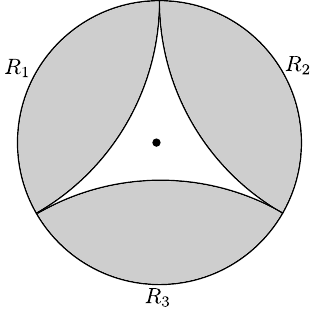}
\caption{The simplest example of holographic quantum error correction: splitting a single AdS boundary into three regions $R_1$, $R_2$, and $R_3$ \cite{Almheiri:2014lwa}.  The white triangular region in the center is an ``island'' in the same sense as the bodies of the octopi in figure \ref{octopi}: it can be accessed from any two of the boundary regions but not from any one.  For example if we compare $R_1$ and $R_2$ to the ``Hawking radiation'' CFTs and $R_3$ to the remaining ``black hole'' CFT, we can view this as a model for an evaporating black hole after the Page time.}\label{qec}
\efig
The ``body'' of the octopus lying between $\gamma$ and $\gamma'$ in figure \ref{octopi} is an example of what was called an ``island'' in \cite{Almheiri:2019hni}.  We emphasize, however, that this island \textit{must} be thought of as a subregion of the bulk theory.  It is not correct to say that it is entangled with the ``Hawking radiation'' CFTs, and it is also not correct to say that after the Page time it is part of those CFTs (the former appears in \cite{Almheiri:2019hni} and we've heard the latter as well).  The set of bulk states consisting of perturbations of the state in our model at any particular time maps into the CFT Hilbert space via an isometric embedding, so the island and the Hawking radiation are subsystems of different Hilbert spaces: one shouldn't discuss entanglement between them.  There is however a natural language of how subsystems of the domain and range of an isometric embedding related: this is the language of quantum error correction (see for example \cite{Harlow:2016vwg} for a review from this point of view).  Indeed properly-construed ``islands''  have been an essential feature of the quantum error-correcting interpretation feature of holography from the beginning: see figure \ref{qec} for an illustration.  In particular we emphasize that even after the Page time in our evaporation model, we can easily find a subset of the ``Hawking radiation'' CFTs whose entanglement wedge does not contain the central island but whose union with the ``black hole'' CFT does have an entanglement wedge that contains the island.  This clearly shows that the island should not be thought of as a subsystem of the ``Hawking radiation'' CFTs (or the auxiliary system which is analogous to them in the models of \cite{Almheiri:2019psf,Penington:2019npb,Almheiri:2019hni}). 

\section{A model for information loss}
\bfig
\includegraphics[height=6cm]{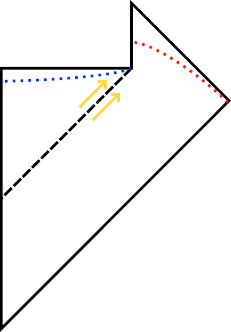}
\caption{The bulk dynamics for information loss: exterior Hawking modes end up in the Hawking radiation (the red dotted line), while interior Hawking modes (and also whatever matter formed the black hole) end up in a baby universe (the blue dotted line).}\label{infoloss}
\efig
We now turn to our second model, which is designed to ``holographically simulate'' what a bulk theory which has information loss would predict.  We state at the outset that we do not actually subscribe to information loss, as we discuss further in the final section, but we believe that this model is nonetheless useful in clarifying how to best interpret the results of \cite{Almheiri:2019psf,Penington:2019npb,Almheiri:2019hni}.  Let's first recall the standard picture of information loss, shown in figure \ref{infoloss} (see \cite{Unruh:2017uaw} for a review on information loss).  Hawking modes are entangled across the black hole horizon, with ``exterior'' modes ending up in the Hawking radiation and ``interior'' modes ending up in a baby universe.  Since there is entanglement between these modes, we end up with entanglement between the Hawking radiation and the baby universe; the Hawking radiation by itself is mixed.  

We will model information loss holographically by again introducing many copies of a single holographic CFT, and beginning in a highly excited pure state in one of the copies and the vacuum in all the others.  We then again couple the CFTs together, but now with a different coupling and a different interpretation: each interaction preserves the purity of the initial ``black hole'' CFT, but decreases its energy and creates a thermofield double pair between two of the other copies.  We will interpret one half of this pair as being part of the baby universe, and the other half as being part of the Hawking radiation.  The baby universes need to be modeled as additional CFTs since in a setting where we take information loss seriously, they are additional degrees of freedom which are independent from those outside of the black hole.  We then continue bringing in additional copies and decreasing the energy of the initial state (see figure \ref{infolossworm} for an illustration).\footnote{Whatever information there is about how the black hole was made also ends up in the baby universe, for simplicity we ignore this.} We further require that the sum of the energy in the ``black hole'' and ``Hawking radiation'' CFTs is conserved, since baby universes should carry zero energy to be viable (recall that really they are closed gravitational systems without boundary). Thus the lengths of $\gamma$ and $\gamma'$ continue to be given by \eqref{Lg} and \eqref{Lgp}.  

\bfig
\includegraphics[height=2.7cm]{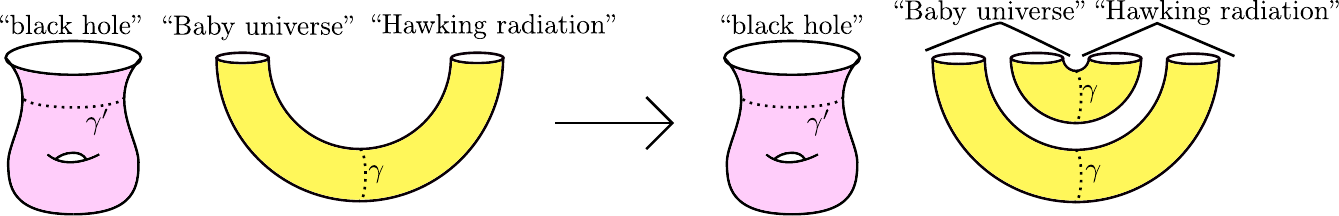}
\caption{The ``black hole evaporation process'' for our second model.  Each step adds an additional copy of the thermofield double state connecting a ``baby universe'' CFT and a ``Hawking radiation'' CFT.  $\gamma'$ is the extremal area surface near the original black hole exterior, and in this model it plays a role analogous to that of the new quantum extremal surface found in \cite{Almheiri:2019psf,Penington:2019npb,Almheiri:2019hni}, but it is null-homologous so it is never the HRT surface for either the ``black hole'' CFT or the ``radiation'' CFTs.  Indeed the HRT surface for the radiation CFTs is always $\gamma$.  We have added a handle to the initial black hole to model whatever pure-state matter we collapsed to form it, in the last step this handle will disappear and the black hole system will be left in vacuum.}\label{infolossworm}
\efig
Eventually the original black hole has radiated down to the vacuum, and we are left just with a set of bipartite wormholes connecting ``baby universe'' CFTs to ``Hawking radiation'' CFTs.  The key point is that now the HRT surface for the Hawking radiation is \textit{always} given by $\gamma$ in figure \ref{infolossworm}: the radiation CFTs are never homologous to the surface $\gamma'$, so there is never a crossover and the von Neumann entropy of the radiation grows linearly until the end, as expected in a theory of information loss.  Moreover since the boundary of the CFT containing the original black hole is always null-homologous, the von Neumann entropy of this black hole is always zero, consistent with the no-hair theorems.  Thus we see that in this model the quantum extremal surface prescription can be perfectly consistent with information loss, provided that we model the bulk in the way that most accurately reflects figure \ref{infoloss}.

\section{Discussion}
We have presented two holographic models for black hole evaporation, one which leads to a final pure state of the Hawking radiation and one which leads to a final mixed state.  In both cases, this conclusion is supported by a quantum extremal surface calculation.  An obvious question is which of these models gives a better analogy to what actually happens to a small black hole in a single holographic CFT evolving without any external modifications of the dynamics.  We believe that the answer is the first model, for the reason that after evaporation in this case the radiation has no choice but to be pure.  More carefully, in a holographic CFT almost all pure states in the microcanonical ensemble at sufficiently low energy are dual to known pure states of the bulk radiation field, so assuming chaotic evolution after sufficiently long times any small black hole will evolve to a pure state which has almost all support on states whose bulk description consists only of soft quanta (see section 6.8 of \cite{Harlow:2014yka} for more discussion of what goes into this argument).  The low-energy Hilbert space of the CFT simply does not have enough states to account for baby universes and/or remnants.\footnote{It is worth noting that our first model does not necessarily give a refutation of the ``firewall'' arguments of \cite{Mathur:2009hf,Almheiri:2012rt,Almheiri:2013hfa,Marolf:2013dba}.  Indeed we could also consider a third model where the spacetime just ends at $\gamma$ and $\gamma'$, and this could also be consistent with unitarity.  Our first model can be viewed as illustrating that some particular microstate can evaporate unitarily while preserving a smooth interior, but to fully address the firewall paradoxes we need to consider what happens in a \textit{typical} microstate.}  

Of course that is not a new argument, it is just the old argument that AdS/CFT solves the information problem in the sense of confirming that the Hawking radiation is pure.  The new issue here is to understand how we should think of the Page curve computed in \cite{Almheiri:2019psf,Penington:2019npb,Almheiri:2019hni}.  The puzzle is the following: their bulk model uses dynamics which are quite close to our second model, but their result for the Page curve is closer to that found in our first model.  To understand how this tension is resolved, one needs to clearly distinguish between two possible calculations:
\bi
\item \textbf{Calculation one}: Couple a holographic CFT in an excited state to an auxiliary system, as first proposed in \cite{Rocha:2008fe}, and evolve using the coupled dynamics.  After a long time, compute the von Neumann entropy of the auxiliary system.  

\item \textbf{Calculation two}: Couple a perturbative semi-classical asymptotically-AdS bulk theory to an auxiliary system via a local coupling at the asymptotic boundary.  After a long time, compute the von Neumann entropy of the auxiliary system. 
\ei 
Calculation one is what we really want to do, but calculation two is what was actually done in \cite{Almheiri:2019psf,Penington:2019npb,Almheiri:2019hni}.  One might hope that the two calculations agree, but they do not.  The reason they do not agree is that they are \textit{not} holographically dual to each other.  Instead we claim that calculation one is dual to a bulk dynamics analogous to our first model, while calculation two is dual to a boundary dynamics analogous to our second model.  The ``baby universe'' CFTs in our second model do not appear in \cite{Almheiri:2019psf,Penington:2019npb,Almheiri:2019hni} because they did not explicitly construct a holographic dual of calculation two.  Had they done so, there would have needed to be a boundary analogue of the degrees of freedom behind the horizon (which are independent from those outside the horizon in perturbative bulk quantum gravity), and in our second model these are our ``baby universe'' CFTs.  The reason \cite{Almheiri:2019psf,Penington:2019npb,Almheiri:2019hni} were able to nonetheless find a Page curve similar to what calculation one would produce is that they allowed themselves to use the quantum extremal surface prescription \textit{assuming} that the auxiliary system is homologous to $\gamma'$, as it is in our first model, even though actually it isn't for the boundary model which is dual to the calculation they actually did. 

What then have we learned?  Our main point is that our first model gives us a situation where a controlled calculation analogous to calculation one from the previous paragraph is possible: we use explicit boundary evolution to carry us through a sequence of CFT states, each of which has a standard and well-known bulk dual, and we use the replica trick to compute the boundary entropy using the HRT formula.  In this model we understand that we are indeed allowed to use the extremal surface $\gamma'$ analogous to the one constructed in \cite{Almheiri:2019psf,Penington:2019npb}, giving us a Page curve that decreases both from the bulk and the boundary point of view, while preserving a smooth interior. We think this is a good model for many aspects of how black holes evaporate in quantum gravity.   Secondly, we've learned (unsurprisingly) from our second model that we should not take perturbative bulk dynamics literally.  Doing so produces information loss, and when a holographic dual is properly constructed it supports this, but it is not compatible with the calculation we actually want to do (calculation one of the previous paragraph).\footnote{We believe we are in agreement with the authors of \cite{Almheiri:2019psf} on this point (we're pretty sure about one of them), we are less sure about \cite{Penington:2019npb}.  Indeed in \cite{Penington:2019npb} there is some suggestion that the final radiation actually is mixed, but in some way that that the CFT state which contains it is still pure.  This does not seem compatible with the argument at the beginning of this conclusion: the low energy CFT states in which the radiation is stored are well-described by bulk effective field theory, as indeed can be confirmed directly from a study of local boundary correlators, and there does not seem to be a way for one to be mixed while the other is pure.}  To phrase it more positively, the analysis of \cite{Almheiri:2019psf,Penington:2019npb,Almheiri:2019hni} has given to us a beautiful new quantum extremal surface to consider, but what we have learned from our two models is that in order to make use of this surface we need to allow for bulk dynamics which go beyond those considered by Hawking.  Indeed the candidate HRT surface $\gamma'$ in our models is precisely analogous to the one constructed in \cite{Almheiri:2019psf,Penington:2019npb,Almheiri:2019hni}, but it was only in our first model that we were allowed to use it in computing the Page curve.  This is all to the good: in order for information to escape from black holes, there \textit{must} be sizable modifications of the bulk dynamics beyond perturbative effective field theory.  The evolution in our first model is such a modification: each time step couples the boundary CFTs together in a non-local way that cannot be interpreted merely as a change of boundary conditions.  The task now seems to be to get a better handle on what this dynamics would look like in a more realistic model.

\paragraph{Acknowledgments}
We'd like to thank Ahmed Almheiri, Raphael Bousso, Venkatesa Chandrasekaran, Adam Levine, Juan Maldacena, Geoff Penington, Pratik Rath, Arvin Shahbazi-Moghaddam, Leonard Susskind, and Ying Zhao for useful discussions. C.A. and D.H. are supported by the US Department of Energy grants DE-SC0018944 and DE-SC0019127, and also the Simons foundation as members of the {\it It from Qubit} collaboration.  D.H. is also supported by the Sloan foundation and the US Department of Defense.  N.E. and D.H are supported by the MIT department of physics.
\bibliographystyle{jhep}
\bibliography{bibliography}

\providecommand{\href}[2]{#2}\begingroup\raggedright\begin{thebibliography}{10}

\bibitem{Hawking:1976ra}
S.~W. Hawking, {\it {Breakdown of Predictability in Gravitational Collapse}},
  {\em Phys. Rev.} {\bf D14} (1976) 2460--2473.

\bibitem{Harlow:2014yka}
D.~Harlow, {\it {Jerusalem Lectures on Black Holes and Quantum Information}},
  {\em Rev. Mod. Phys.} {\bf 88} (2016) 015002,
  [\href{http://arxiv.org/abs/1409.1231}{{\tt arXiv:1409.1231}}].

\bibitem{Marolf:2017jkr}
D.~Marolf, {\it {The Black Hole information problem: past, present, and
  future}},  {\em Rept. Prog. Phys.} {\bf 80} (2017), no.~9 092001,
  [\href{http://arxiv.org/abs/1703.02143}{{\tt arXiv:1703.02143}}].

\bibitem{Almheiri:2019psf}
A.~Almheiri, N.~Engelhardt, D.~Marolf, and H.~Maxfield, {\it {The entropy of
  bulk quantum fields and the entanglement wedge of an evaporating black
  hole}},  \href{http://arxiv.org/abs/1905.08762}{{\tt arXiv:1905.08762}}.

\bibitem{Penington:2019npb}
G.~Penington, {\it {Entanglement Wedge Reconstruction and the Information
  Paradox}},  \href{http://arxiv.org/abs/1905.08255}{{\tt arXiv:1905.08255}}.

\bibitem{Almheiri:2019hni}
A.~Almheiri, R.~Mahajan, J.~Maldacena, and Y.~Zhao, {\it {The Page curve of
  Hawking radiation from semiclassical geometry}},
  \href{http://arxiv.org/abs/1908.10996}{{\tt arXiv:1908.10996}}.

\bibitem{Rocha:2008fe}
J.~V. Rocha, {\it {Evaporation of large black holes in AdS: Coupling to the
  evaporon}},  {\em JHEP} {\bf 08} (2008) 075,
  [\href{http://arxiv.org/abs/0804.0055}{{\tt arXiv:0804.0055}}].

\bibitem{Engelhardt:2014gca}
N.~Engelhardt and A.~C. Wall, {\it {Quantum Extremal Surfaces: Holographic
  Entanglement Entropy beyond the Classical Regime}},  {\em JHEP} {\bf 01}
  (2015) 073, [\href{http://arxiv.org/abs/1408.3203}{{\tt arXiv:1408.3203}}].

\bibitem{Czech:2012bh}
B.~Czech, J.~L. Karczmarek, F.~Nogueira, and M.~Van~Raamsdonk, {\it {The
  Gravity Dual of a Density Matrix}},  {\em Class. Quant. Grav.} {\bf 29}
  (2012) 155009, [\href{http://arxiv.org/abs/1204.1330}{{\tt
  arXiv:1204.1330}}].

\bibitem{Wall:2012uf}
A.~C. Wall, {\it {Maximin Surfaces, and the Strong Subadditivity of the
  Covariant Holographic Entanglement Entropy}},  {\em Class. Quant. Grav.} {\bf
  31} (2014), no.~22 225007, [\href{http://arxiv.org/abs/1211.3494}{{\tt
  arXiv:1211.3494}}].

\bibitem{Headrick:2014cta}
M.~Headrick, V.~E. Hubeny, A.~Lawrence, and M.~Rangamani, {\it {Causality \&
  holographic entanglement entropy}},  {\em JHEP} {\bf 12} (2014) 162,
  [\href{http://arxiv.org/abs/1408.6300}{{\tt arXiv:1408.6300}}].

\bibitem{Jafferis:2015del}
D.~L. Jafferis, A.~Lewkowycz, J.~Maldacena, and S.~J. Suh, {\it {Relative
  entropy equals bulk relative entropy}},  {\em JHEP} {\bf 06} (2016) 004,
  [\href{http://arxiv.org/abs/1512.06431}{{\tt arXiv:1512.06431}}].

\bibitem{Dong:2016eik}
X.~Dong, D.~Harlow, and A.~C. Wall, {\it {Reconstruction of Bulk Operators
  within the Entanglement Wedge in Gauge-Gravity Duality}},  {\em Phys. Rev.
  Lett.} {\bf 117} (2016), no.~2 021601,
  [\href{http://arxiv.org/abs/1601.05416}{{\tt arXiv:1601.05416}}].

\bibitem{Harlow:2016vwg}
D.~Harlow, {\it {The Ryu-Takayanagi Formula from Quantum Error Correction}},
  {\em Commun. Math. Phys.} {\bf 354} (2017), no.~3 865--912,
  [\href{http://arxiv.org/abs/1607.03901}{{\tt arXiv:1607.03901}}].

\bibitem{Ryu:2006bv}
S.~Ryu and T.~Takayanagi, {\it {Holographic derivation of entanglement entropy
  from AdS/CFT}},  {\em Phys. Rev. Lett.} {\bf 96} (2006) 181602,
  [\href{http://arxiv.org/abs/hep-th/0603001}{{\tt hep-th/0603001}}].

\bibitem{Hubeny:2007xt}
V.~E. Hubeny, M.~Rangamani, and T.~Takayanagi, {\it {A Covariant holographic
  entanglement entropy proposal}},  {\em JHEP} {\bf 07} (2007) 062,
  [\href{http://arxiv.org/abs/0705.0016}{{\tt arXiv:0705.0016}}].

\bibitem{Faulkner:2013ana}
T.~Faulkner, A.~Lewkowycz, and J.~Maldacena, {\it {Quantum corrections to
  holographic entanglement entropy}},  {\em JHEP} {\bf 11} (2013) 074,
  [\href{http://arxiv.org/abs/1307.2892}{{\tt arXiv:1307.2892}}].

\bibitem{Hayden:2018khn}
P.~Hayden and G.~Penington, {\it {Learning the Alpha-bits of Black Holes}},
  \href{http://arxiv.org/abs/1807.06041}{{\tt arXiv:1807.06041}}.

\bibitem{Akers:2019wxj}
C.~Akers, A.~Levine, and S.~Leichenauer, {\it {Large Breakdowns of Entanglement
  Wedge Reconstruction}},  \href{http://arxiv.org/abs/1908.03975}{{\tt
  arXiv:1908.03975}}.

\bibitem{Krasnov:2000zq}
K.~Krasnov, {\it {Holography and Riemann surfaces}},  {\em Adv. Theor. Math.
  Phys.} {\bf 4} (2000) 929--979,
  [\href{http://arxiv.org/abs/hep-th/0005106}{{\tt hep-th/0005106}}].

\bibitem{Skenderis:2009ju}
K.~Skenderis and B.~C. van Rees, {\it {Holography and wormholes in 2+1
  dimensions}},  {\em Commun. Math. Phys.} {\bf 301} (2011) 583--626,
  [\href{http://arxiv.org/abs/0912.2090}{{\tt arXiv:0912.2090}}].

\bibitem{Balasubramanian:2014hda}
V.~Balasubramanian, P.~Hayden, A.~Maloney, D.~Marolf, and S.~F. Ross, {\it
  {Multiboundary Wormholes and Holographic Entanglement}},  {\em Class. Quant.
  Grav.} {\bf 31} (2014) 185015, [\href{http://arxiv.org/abs/1406.2663}{{\tt
  arXiv:1406.2663}}].

\bibitem{Harlow:2018tng}
D.~Harlow and H.~Ooguri, {\it {Symmetries in quantum field theory and quantum
  gravity}},  \href{http://arxiv.org/abs/1810.05338}{{\tt arXiv:1810.05338}}.

\bibitem{Hayden:2007cs}
P.~Hayden and J.~Preskill, {\it {Black holes as mirrors: Quantum information in
  random subsystems}},  {\em JHEP} {\bf 09} (2007) 120,
  [\href{http://arxiv.org/abs/0708.4025}{{\tt arXiv:0708.4025}}].

\bibitem{Lewkowycz:2013nqa}
A.~Lewkowycz and J.~Maldacena, {\it {Generalized gravitational entropy}},  {\em
  JHEP} {\bf 08} (2013) 090, [\href{http://arxiv.org/abs/1304.4926}{{\tt
  arXiv:1304.4926}}].

\bibitem{VanRaamsdonk:2010pw}
M.~Van~Raamsdonk, {\it {Building up spacetime with quantum entanglement}},
  {\em Gen. Rel. Grav.} {\bf 42} (2010) 2323--2329,
  [\href{http://arxiv.org/abs/1005.3035}{{\tt arXiv:1005.3035}}]. [Int. J. Mod.
  Phys.D19,2429(2010)].

\bibitem{VanRaamsdonk:2013sza}
M.~Van~Raamsdonk, {\it {Evaporating Firewalls}},  {\em JHEP} {\bf 11} (2014)
  038, [\href{http://arxiv.org/abs/1307.1796}{{\tt arXiv:1307.1796}}].

\bibitem{Maldacena:2013xja}
J.~Maldacena and L.~Susskind, {\it {Cool horizons for entangled black holes}},
  {\em Fortsch. Phys.} {\bf 61} (2013) 781--811,
  [\href{http://arxiv.org/abs/1306.0533}{{\tt arXiv:1306.0533}}].

\bibitem{Mathur:2009hf}
S.~D. Mathur, {\it {The Information paradox: A Pedagogical introduction}},
  {\em Class. Quant. Grav.} {\bf 26} (2009) 224001,
  [\href{http://arxiv.org/abs/0909.1038}{{\tt arXiv:0909.1038}}].

\bibitem{Almheiri:2012rt}
A.~Almheiri, D.~Marolf, J.~Polchinski, and J.~Sully, {\it {Black Holes:
  Complementarity or Firewalls?}},  {\em JHEP} {\bf 02} (2013) 062,
  [\href{http://arxiv.org/abs/1207.3123}{{\tt arXiv:1207.3123}}].

\bibitem{Almheiri:2013hfa}
A.~Almheiri, D.~Marolf, J.~Polchinski, D.~Stanford, and J.~Sully, {\it {An
  Apologia for Firewalls}},  {\em JHEP} {\bf 09} (2013) 018,
  [\href{http://arxiv.org/abs/1304.6483}{{\tt arXiv:1304.6483}}].

\bibitem{Marolf:2013dba}
D.~Marolf and J.~Polchinski, {\it {Gauge/Gravity Duality and the Black Hole
  Interior}},  {\em Phys. Rev. Lett.} {\bf 111} (2013) 171301,
  [\href{http://arxiv.org/abs/1307.4706}{{\tt arXiv:1307.4706}}].

\bibitem{Almheiri:2014lwa}
A.~Almheiri, X.~Dong, and D.~Harlow, {\it {Bulk Locality and Quantum Error
  Correction in AdS/CFT}},  {\em JHEP} {\bf 04} (2015) 163,
  [\href{http://arxiv.org/abs/1411.7041}{{\tt arXiv:1411.7041}}].

\bibitem{Unruh:2017uaw}
W.~G. Unruh and R.~M. Wald, {\it {Information Loss}},  {\em Rept. Prog. Phys.}
  {\bf 80} (2017), no.~9 092002, [\href{http://arxiv.org/abs/1703.02140}{{\tt
  arXiv:1703.02140}}].

\end{thebibliography}\endgroup
\end{document}